\documentclass[twocolumn,showpacs,preprintnumbers,amsmath,amssymb,prl]{revtex4}

\usepackage[]{graphicx}
\usepackage{amsmath}

\begin{document}

\title[]{Non-Arrhenius Behavior of Secondary Relaxation in Supercooled Liquids}

\author{Takuya Fujima}
\email{takuya@exp.t.u-tokyo.ac.jp}

\author{Hiroshi Frusawa}
\email{furu@exp.t.u-tokyo.ac.jp}

\affiliation{Department of Applied Physics, School of Engineering,
University of Tokyo, 7-3-1 Hongo, Bunkyo-ku, Tokyo 113-8656, Japan}

\author{Kohzo Ito}

\affiliation{Department of Advanced Materials Science, School of Frontier Sciences, University of Tokyo, 7-3-1 Hongo, Bunkyo-ku, Tokyo 113-8656, Japan}

\date{\today}

\begin{abstract}
Dielectric relaxation spectroscopy ($1\,\mbox{\rm{Hz}}$ -- $20\,\mbox{\rm{GHz}}$) has been performed on supercooled glass-formers from the temperature of glass transition ($T_{\mbox{\tiny g}}$) up to that of melting. Precise measurements particularly in the frequencies of MHz--order have revealed that the temperature dependences of secondary $\beta$--relaxation times deviate from the Arrhenius relation in well above $T_{\mbox{\tiny g}}$. Consequently, our results indicate that the $\beta$--process merges into the primary $\alpha$--mode around the melting temperature, and not at the dynamical transition point $T\approx 1.2 T_{\mbox{\tiny g}}$.
\end{abstract}


\maketitle

In recent years, much of the focus on glassy dynamics has been shifting to a considerably higher temperature than $T_{\mbox{\tiny g}}$, the glass-transition one~\cite{rev_angell,rev_donth,rev_idmrcs,rev_sillescu}. The topical temperature is located around $T_{\mbox{\tiny D}}\equiv 1.2 T_{\mbox{\tiny g}}$, where the dynamics of supercooled liquids has been found to change fairly. So far, there have been observed the following phenomena:

  \begin{itemize}
     \item[(i)] R\"{o}ssler scaling reveals that, when cooled, the Debye-Stokes-Einstein relation becomes invalid around $T_{\mbox{\tiny D}}$~\cite{capacci,rossler_prl}. This indicates a change of diffusion mechanism there.
     \item[(ii)] Stickel analysis~\cite{stickel} clarifies that temperature dependence of viscosity changes around $T_{\mbox{\tiny D}}$. Therefore, in order to fit the primary $\alpha$--relaxation time $\tau_{\alpha}$ using the Vogel-Fulcher-Tamman (VFT) relation,
					\begin{eqnarray}
						\tau_{\alpha} =\tau_{\mbox{\tiny 0}}\exp \left( {{C \over {T-T_{\mbox{\tiny 0}}}}} \right),
					\end{eqnarray}
the coefficients ($\tau_{\mbox{\tiny 0}}, C, T_{\mbox{\tiny 0}}$) have to vary at $T_{\mbox{\tiny B}}\approx T_{\mbox{\tiny D}}$. This suggests that the mechanism of slow structural-relaxation makes some alternation there.
     \item[(iii)] Johari-Goldstein type $\beta$--process (secondary process in the context of dielectric relaxation)~\cite{jg,faivre,lunkenheimer,kremer,nozaki,kudlik} merges into the $\alpha$--relaxation around $T_D$, extrapolating the Arrhenius-type temperature dependence below $T_{\mbox{\tiny g}}$ of the $\beta$--relaxation times~\cite{rossler_prl,donth,shroter,richert,angell_10q,richter,rault}.
  \end{itemize}

Theoretically, the characteristic temperature $T_{\mbox{\tiny D}}$ is thought to be comparable to $T_{\mbox{\tiny C}}$ where the idealized Mode Coupling Theory (MCT) predicts a dynamical phase transition~\cite{mct}. Indeed, the above first two phenomena (i, ii) can be regarded as indicators of the dynamical transition at $T_{\mbox{\tiny C}}$. However, the MCT is irrelevant to the third one (iii), the bifurcation of $\alpha, \beta$--modes; even the existence of the Johari-Goldstein type $\beta$--process cannot be derived.

Furthermore, from experimental aspects, while the dynamical transition phenomena (i, ii) have been confirmed from either R\"{o}ssler or Stickel plot definitely, the bifurcation of (iii) is inferred from the extrapolation. Actually, however, it remains an open problem as to whether the Arrhenius behavior of the $\beta$--process persists in higher temperatures near $T_{\mbox{\tiny D}}$: the $T_{\mbox{\tiny D}}$--decoupling of $\alpha, \beta$--relaxations is not conclusive.

This letter thus aims to investigate the secondary $\beta$--mode in high temperatures well above $T_{\mbox{\tiny g}}$, by carrying out precisely the broad-band measurements of dielectric relaxation. Our main result is the following: as will be seen in Figures 2 and 5, the $\beta$--relaxation times deviate from the Arrhenius relation, indicating that {\itshape the $\beta$--relaxation merges into the $\alpha$--mode not at $T_{\mbox{\tiny D}}$ but around the Arrhenius-VFT crossover temperature $T_{\mbox{\tiny A}}$} (generally close to the melting one) where the temperature dependence of the $\alpha$--process changes between an Arrhenius type and a VFT one.

In order to trace the $\beta$--mode in wide temperature range from $T_{\mbox{\tiny g}}$ to $T_{\mbox{\tiny A}}$, we set up a broad-band dielectric relaxation measurement system that the following three apparatuses can work simultaneously: Sorlatron1260 impedance analyzer with parallel-plate electrodes (1 Hz -- 32 MHz), HP4191A reflectance analyzer with coaxially-cylindrical electrodes (1 MHz -- 1 GHz), and HP54750A with HP54754A Time Domain Reflectometry (TDR) system with flat-end-cable-type electrodes (10 MHz -- 20 GHz). The three sets of electrodes were installed in the same sample bath for their identical temperature control.

\begin{figure*}
\includegraphics[scale=1]{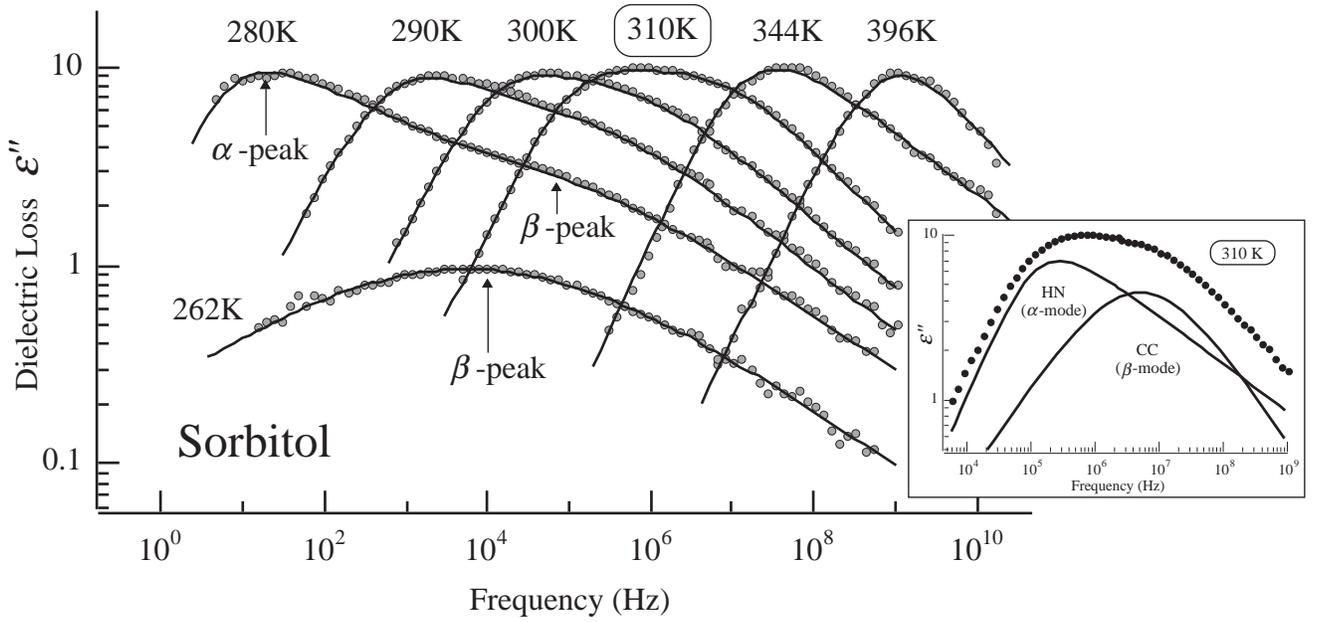}
\caption{\label{fig:wide}Dielectric loss spectra of sorbitol at various temperatures.  Circles represent the experimental data, and the solid curves are the best fitting results by eq. (2). In the inset, two relaxations for the data of 310 K are specified.}
\end{figure*}

As a typical glass-forming sample with a large strength of the $\beta$--relaxation, we employed sorbitol (CH$_2$OH-(CHOH)$_4$-CH$_2$OH, $T_{\mbox{\tiny m}} = 383$ K, $T_{\mbox{\tiny g}}= 264$ K) purchased from Nacalai Tesque Co.. In preparation, we heated the sample at 400 K for 20 minutes in vacuum for complete melting and purification. The obtained simple-liquid-state sample loaded in the sample bath was quickly quenched into deeply supercooled liquid near $T_{\mbox{\tiny g}}$ at the cooling rate of 56.7 K/min. We then started measurements isothermally at specified temperatures under an accuracy to within 0.1 K. These series of procedures were done in dry-nitrogen atmosphere to prevent the samples from absorbing moisture.

Figure 1 shows the dielectric loss spectra of sorbitol at various temperatures, exhibiting an obvious $\beta$--mode in lower temperatures and its overlap with the $\alpha$ --mode as temperature increases. It should be noticed that two modes of $\alpha, \beta$--relaxations can be discriminated even around $T_{\mbox{\tiny D}} = 1.2 T_{\mbox{\tiny g}}$ (see $310$K=1.17$T_{\mbox{\tiny g}}$ in Fig. 1, for example). That is, only observing dielectric spectra reveals that {\itshape the $\beta$--relaxation does not merge into the $\alpha$--mode around $T_{\mbox{\tiny B}}\approx T_{\mbox{\tiny D}}$}.

\begin{figure}
\includegraphics[scale=1]{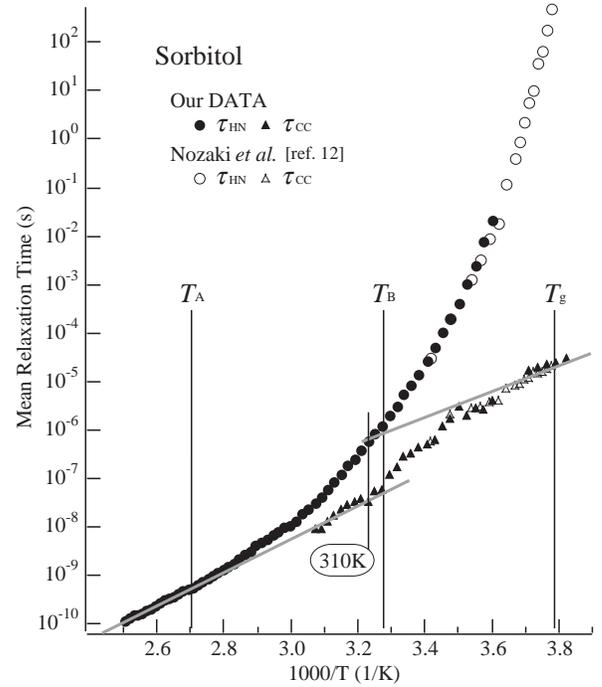}
\caption{Temperature dependence of the $\alpha, \beta$--relaxation times of sorbitol, which exhibits good consistency with low temperature data reported so far~\cite{nozaki}.}
\end{figure}

In the line-shape analysis of these spectra by the method of least squares, we used the Havriliak-Negami (HN) function for fitting the $\alpha$--process and the Cole-Cole (CC) function for the $\beta$--mode. These are empirical relaxation functions including broadness and skewness of the relaxation time distributions: 
\begin{eqnarray}
\varepsilon ^*=\varepsilon _\infty +{{\Delta \varepsilon _{\mbox{\tiny {HN}}}} \over {\left( {1+\left( {i\omega \tau _{\mbox{\tiny {HN}}}}\right)^{\alpha _{\mbox{\tiny {HN}}}}} \right)^{\beta _{\mbox{\tiny {HN}}}}}}+{{\Delta \varepsilon _{\mbox{\tiny {CC}}}} \over {1+\left({i\omega \tau _{\mbox{\tiny {CC}}}} \right)^{\alpha _{\mbox{\tiny {CC}}}}}}.
\end{eqnarray}
Here ${\Delta \varepsilon _{\mbox{\tiny {HN}}}}$ and ${\Delta \varepsilon_{\mbox{\tiny {CC}}}}$ are the dielectric increments, ${\tau _{\mbox{\tiny {HN}}}}$ and ${\tau _{\mbox{\tiny {CC}}}}$ are the relaxation times, ${\alpha _{\mbox{\tiny {HN}}}}$ and ${\alpha _{\mbox{\tiny {CC}}}}$ are the parameters of distribution broadness of ${\tau _{\mbox{\tiny {HN}}}}$ and ${\tau _{\mbox{\tiny {CC}}}}$, respectively, ${\beta _{\mbox{\tiny {HN}}}}$ is a parameter reflecting the skewness of ${\tau _{\mbox{\tiny {HN}}}}$ distribution, ${\varepsilon _\infty}$ is the high-frequency limit of ${\varepsilon ^*}$, and ${\omega}$ is the angular frequency.

Let us first show in Figure 2 the temperature dependences of the $\alpha,\beta$--relaxations times obtained from the above analysis. Our data are smoothly connected to the results reported so far~\cite{nozaki}. From the Stickel analysis on the $\alpha$--relaxation, we get the Arrhenius-VFT crossover temperature and the VFT-VFT one as $T_{\mbox{\tiny A}} = 370$ K and $T_{\mbox{\tiny B}} = 305$ K, respectively. From the $\beta$--process, on the other hand, two features are found:

  \begin{itemize}
    \item With heating, the Arrhenius relation below $T_{\mbox{\tiny g}}$ does not persist to $T_{\mbox{\tiny B}}\approx T_{\mbox{\tiny D}}$. The $\beta$--relaxation times start to decrease with larger activation energy just before $T_{\mbox{\tiny B}}$.
    \item Above $T_{\mbox{\tiny B}}$, the activation energy of the $\beta$--mode becomes nearly equal to that of the $\alpha$--process above $T_{\mbox{\tiny A}}$. Consequently, the $\beta$--process appears to approach smoothly the $\alpha$--process around $T_{\mbox{\tiny A}}$~\cite{margulies}.
    \end{itemize}
  
Indeed, as guided in Fig. 2, extrapolating the Arrhenius dependence below $T_{\mbox{\tiny g}}$ leads to the $\alpha$-$\beta$ bifurcation around $T_{\mbox{\tiny B}}\approx T_{\mbox{\tiny D}}$ in accordance with previous results. Our precise measurements, however, reveal the invalidity of such analysis.

Other fitting parameters also support the $T_{\mbox{\tiny A}}$--merging of $\alpha,\beta$-modes as seen in Figures 3, the plot of the temperature versus dielectric increments ($\Delta\epsilon_{\mbox{\tiny {HN}}}, \Delta\epsilon_{\mbox{\tiny {CC}}}$) and fitting parameters ($\alpha_{\mbox{\tiny {HN}}}, \alpha_{\mbox{\tiny {CC}}}, \beta_{\mbox{\tiny {HN}}}$). Especially look at the behavior of $\beta$--process, showing that both the dielectric increment $\Delta\epsilon_{\mbox{\tiny {CC}}}$ and the broadness parameter $\alpha_{\mbox{\tiny {CC}}}$ increase gradually with heating and approach the value of the $\alpha$--relaxation. To be noted, $\alpha_{\mbox{\tiny {CC}}}$ appears to merge into that of the $\alpha$--mode in $2.8\leq 1000/T\leq 3$, similarly to the result of relaxation times.

\begin{figure}
\includegraphics[scale=1]{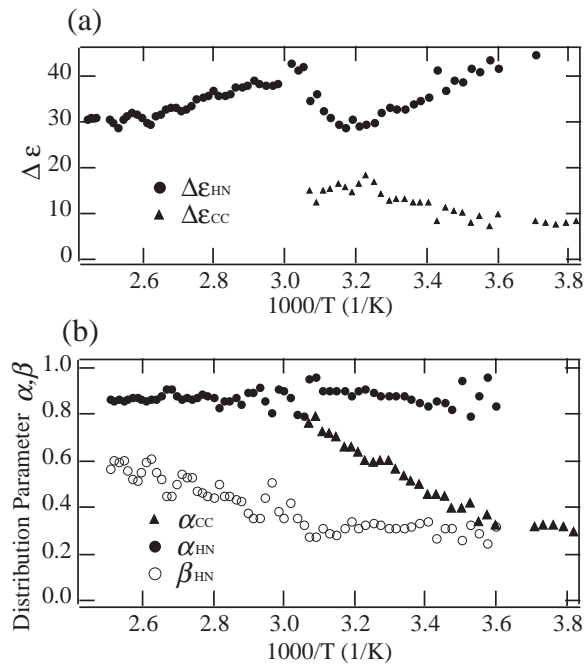}
\caption{Temperature dependences, for sorbitol, of (a) dielectric increments $\Delta\epsilon_{\mbox{\tiny {HN}}}, \Delta\epsilon_{\mbox{\tiny {CC}}}$ and (b) broadness parameters $\alpha_{\mbox{\tiny {HN}}}, \alpha_{\mbox{\tiny {CC}}}$ and skewness parameter $\beta_{\mbox{\tiny {HN}}}$.}
\end{figure}

We have further checked the non-Arrhenius behavior for another material of entirely different-molecular structure: m-fluoroaniline (FAN) (NH$_2$-C$_6$H$_4$-F, $T_{\mbox{\tiny m}} = 238$ K, $T_{\mbox{\tiny g}} = 169$ K). Figure 4 shows the temperature dependences of $\alpha, \beta$--relaxations times, which have good continuity with previous ones as before~\cite{kudlik}. Remarkably, the $\beta$--process of FAN shows the rapid change of relaxation time around $T_{\mbox{\tiny B}}$ and smooth approach to the $\alpha$--mode around $T_{\mbox{\tiny A}}$, identically to that of sorbitol.

We would like to thus consider the molecular mechanism behind the non-Arrhenius behavior of the $\beta$--mode now revealed, classifying the temperatures into three regimes: $T\leq T_{\mbox{\tiny g}}$ (region I), $T_{\mbox{\tiny g}}\leq T\leq T_{\mbox{\tiny B}}$  (region II) and $T_{\mbox{\tiny B}}\leq T\leq T_{\mbox{\tiny A}}$ (region III). Naturally, our main concern will be given to region II.

{\it Region I} --- Most of the investigations on the $\beta$--mode have been done here. Among them, recent NMR--study provides us with the details of the $\beta$--relaxation: the Arrhenius-type temperature dependence in this region follows from some short-angle libration of molecules confined by effectively frozen surroundings (i.e., in "cages")~\cite{cage_ww,rossler_nmr}.

\begin{figure}
\includegraphics[scale=1]{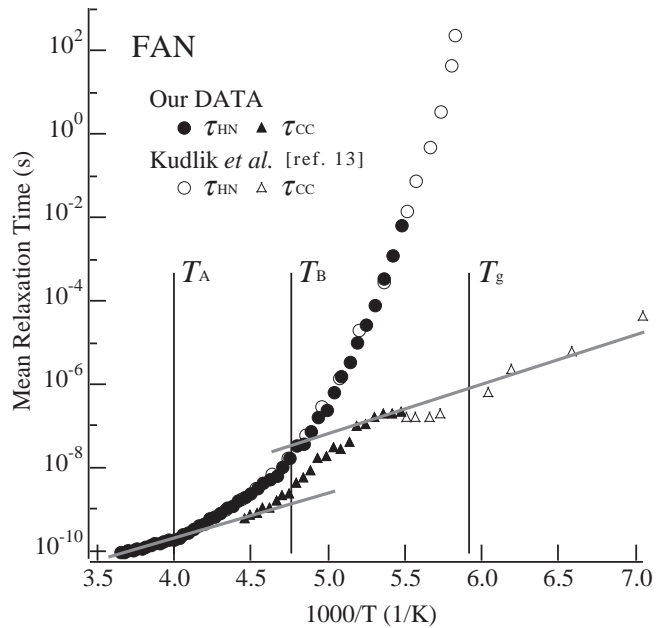}
\caption{Temperature dependence of the $\alpha, \beta$--relaxation times of m-fluoroaniline. The temperature behavior is similar to that of sorbitol displayed in Fig. 2.}
\end{figure}

{\it Region II} --- Let us focus here the steep change of the $\beta$-relaxation times, i.e., the larger activation energy than those in other regions. A possible scenario to explain this is the following.

We rely on the theoretical suggestion that "cages" start to melt around $T_{\mbox{\tiny D}}$~\cite{mct}. With this, the $T_{\mbox{\tiny D}}$--change of the $\alpha$-process [i.e., (ii) described at the beginning] has been ascribed. Now suppose that the half-melt of cages reflects also the $\beta$-process: we assume that the loosening of confinement around $T_{\mbox{\tiny D}}$ enables some molecules to move, within the $\beta$-relaxation time, from one cage to another with scratching surroundings. It is then plausible that the $\beta$-relaxation involving such $\alpha$-like-process requires more activation energy than that of the libration in Region I.

{\it Region III} --- In the highest temperature region, the $\beta$-relaxation merges into the $\alpha$--process around $T_{\mbox{\tiny A}}$ with similar activation energies. This suggests that, due to further softening of cages, the $\beta$--rotation becomes free as in simple liquids before the $\alpha$--mode does.

In summary, we performed broad-band dielectric spectroscopy (1 Hz - 20 GHz) on supercooled sorbitol and m-fluoroaniline with focusing on the behavior of the $\beta$--mode in well above $T_{\mbox{\tiny g}}$ region. From obtained results, it has been revealed that the $\beta$--relaxation times rapidly decrease around the VFT-VFT crossover temperature $T_{\mbox{\tiny B}}\approx T_{\mbox{\tiny D}}$ with larger activation energy than that below $T_{\mbox{\tiny g}}$. Moreover, above $T_{\mbox{\tiny B}}$, the $\beta$--mode shifts into the same temperature dependence as that of the simple-liquid state to merge into the $\alpha$--mode around the Arrhenius-VFT crossover temperature $T_{\mbox{\tiny A}}$. These results suggest that cage-melting around $T_{\mbox{\tiny D}}$ changes the dynamics of the $\beta$--process as well as the $\alpha$--mode. The non-Arrhenius behavior of secondary relaxation, common to two materials of different molecular structures, is expected to be universal for supercooled liquids, though wider materials should be checked carefully in future.

We thank N. Matsuda for his great help on the use of an apparatus. This work is supported by the JST-CREST. One of the authors (T. F.) also acknowledges the financial support from the JSPS Research Fellowship for Young Scientists.


\end{document}